\documentclass[%
 reprint,
groupedaddress,
 amsmath,amssymb,
 aps,
prl,
]{revtex4-2}

\usepackage{graphicx}
\usepackage{dcolumn}
\usepackage{bm}
\usepackage{filecontents}
\usepackage{xcolor}
\usepackage{ulem}
\def\e{\begin{equation}}
\def\f{\end{equation}}

\def\=#1{\overline{\overline #1}}

\def\.{\cdot}

\def\_#1{{\bf #1\mit}}

\def\@#1{_{\rm #1}}

\begin{document}

\preprint{APS/123-QED}

\title{Coherent Retroreflective Metasurfaces
}

\author{F.~S.~Cuesta}
\author{G.~A.~Ptitcyn}
\author{M.~S.~Mirmoosa}
\author{S.~A.~Tretyakov}
\affiliation{Department of Electronics and Nanoengineering, Aalto University, P.O.~Box 15500, FI-00076 Aalto, Finland}

\date{\today}

\begin{abstract}
Inhomogeneous metasurfaces have shown possibilities for  unprecedented control of wave propagation and scattering. While it is conventional to shine a single incident plane wave from one side of these metastructures, illuminating by several waves simultaneously from both sides may enhance possibilities to control scattered waves, which results in additional functionalities and novel applications. Here, we unveil how using coherent plane-wave illumination of a properly designed inhomogeneous metasurface sheet it is possible to realize controllable retroreflection. We call these metasurfaces as ``coherent retroreflectors" and explain the method for realizing them both in theory and practice. We show that coherent retroreflectors can be used for filtering undesired modes and creation of field-localization regions in waveguides.  The latter application is in resemblance to bound states in the radiation continuum.  
\end{abstract}

\maketitle

Metasurfaces, as optically thin layers structured at the subwavelength scale~(e.g., \cite{GLYBOVSKI20161,metasurfaces111,metasurfaces222}), introduce discontinuities for tangential components of electric and magnetic fields. We consider nonmagnetic layers with negligible thickness, which support exclusively electric surface currents. This condition limits possible functionalities as these  metasurfaces only create a discontinuity of the magnetic field. For instance, homogeneous thin sheets of any material can absorb maximum 50\% of the incident power of electromagnetic waves. However, this limitation is removed if both sides of the sheet are coherently illuminated, which can be achieved by positioning a reflector behind the sheet~\cite{absorbers_review} or by illuminating the other side with a coherent wave~\cite{Alu_coherent_review,Li_2015} created by a splitter or an independent phase-locked generator. Similar observation can be made for fully reflective sheets. A single uniform sheet of negligible thickness can fully reflect incident waves only if it is perfectly conducting. Thus, there is no possibility to control reflection phase. This limitation can be also removed by adding a reflector behind the sheet, creating so-called high-impedance surfaces~\cite{Sievenpiper_1999} or illuminating the two sides of the sheet by two coherent waves~\cite{Zheludev_coherent_review}. 

Despite these possibilities, homogeneous coherent metasurfaces cannot change the direction of propagation of incident waves. In contrast, optically thin sheets whose properties vary over the surface have this important capacity. In consequence, they can perform a multitude of functions, such as anomalous reflection, anomalous transmission, power splitting, focusing, surface-wave control, and so forth~(e.g.,~\cite{GLYBOVSKI20161}). Nevertheless, similarly to uniform sheets, the functionalities of inhomogeneous sheets are also limited because the tangential electric field component still remains continuous  across the sheet. 
Following an analogy with uniform coherent metasurfaces, we expect  that coherent illuminations of surface-inhomogeneous sheets can dramatically widen the scope of controllable electromagnetic effects. 

\begin{figure}[t!]
\centering
\includegraphics[width=1\linewidth]{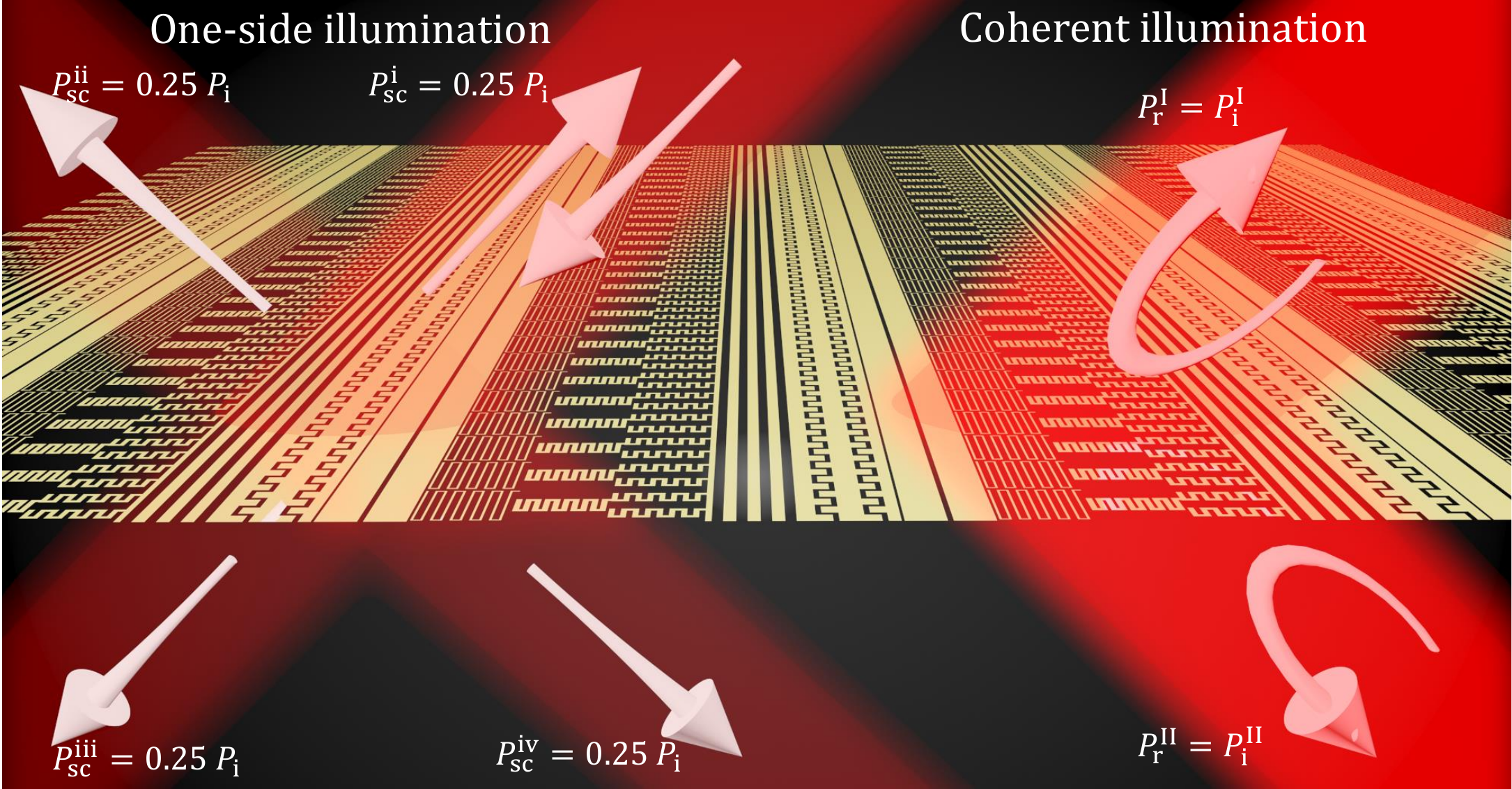}
\caption{A coherent retroreflective metasurface which is illuminated by two incident plane waves, from top and bottom, and reflects those incident waves fully in the same directions, without any parasitic additional modes.}
\label{fig:fig_concept}
\end{figure}

In this letter, we introduce surface-inhomogeneous coherent metasurfaces which can be defined as the surface-varying sheets with extended properties under coherent illumination from both sides. In particular, this letter introduces and studies coherent metasurfaces designed for total retroreflection, as illustrated in Fig.~\ref{fig:fig_concept}.

Retroreflectors that are capable of reflecting the electromagnetic energy back to the source are essential for different important applications \cite{lasertracking,opticaltags,communication1,communication2,communication3,wirelesssensing,remotesensing,Cuesta_2020_retroboundary}. 
Similarly to the case of uniform sheets of negligible optical thickness, the performance of non-uniform sheets designed for retroreflection is fundamentally limited: they cannot reflect more than $25\%$ of the incident power into the desired direction (see Supplementary Information). To dramatically enhance the efficiency, one can propound an  elegant solution to use a {\it coherent} retroreflective sheet (as presented in Fig.~\ref{fig:fig_concept}) which can retroreflect completely the power from both sources.

\begin{figure*}[t!]
\centering
\includegraphics[width=\textwidth]{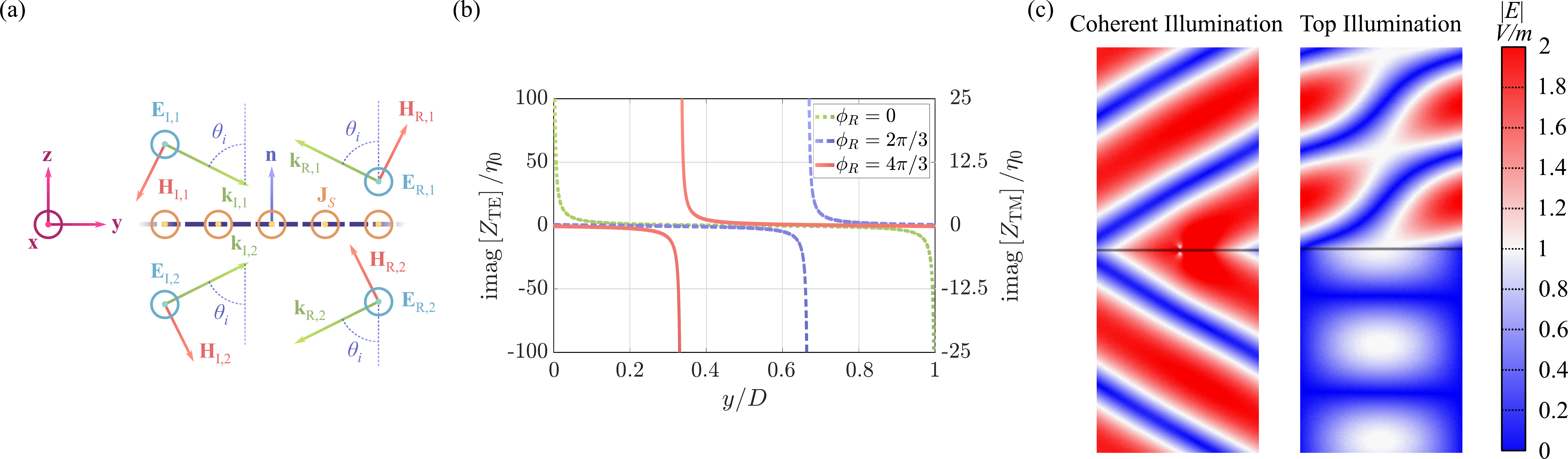}
\caption{\label{fig:teo_group} (a) Schematic representation of a coherent retroreflector for TE waves. (b) Normalized reactance profile for TE- and TM-polarized coherent retroreflectors. The impedance for the two polarizations differ by the scale factor $\cos(\theta_i)^2$ ($\theta_i=60^{\circ}$ for this illustration), using different values of $\phi_R$. (c) The coherent retroreflective sheet exhibits total retroreflection when it is illuminated from both sides; however, additional scattering is produced when it is illuminated from only one side ($\theta_i=60^{\circ}$, $\phi_R=0^{\circ}$ and $f=10$ GHz for this illustration).}
\end{figure*}

We explain how to design coherent retroreflective metasurfaces for both transverse-electric (TE) and transverse-magnetic (TM) linearly polarized waves. Furthermore, we  discuss how the proposed retroreflective metasurface can be effectively used to suppress unwanted waveguide modes and localize high-amplitude fields in unbounded waveguiding structures. In addition, we note that it is possible to replace  conventional retroreflectors~\cite{Acoleyen_2011,Nagayama_2016,Wong_2018,Jia_2018,Wang_2018_asymmetry,Wang_2020,Liu_2020,Diaz-Rubio_2020,Cuesta_2020_retroboundary} with extremely light and thin coherent  structures. Moreover, the coherent nature of retroreflection of the proposed structures also allows tunability, for instance,  by turning the second wave on or off.

Let us consider a metasurface sheet of negligible thickness as an interface  between two vacuum regions, as shown in Fig.~\ref{fig:teo_group}(a). We assume that the sheet supports only electric surface current, with the surface current density $\_J_{s}$, and it is illuminated from both sides by plane waves with the incidence angle $\theta_i$. Such metasurface can be characterized by the boundary condition \mbox{ $Z_{s} \_J_{s}=\left(\_E\@{\tau,1}+\_E\@{\tau,2}\right)/2$}, where $\_E\@{\tau,1-2}$ ($\_H\@{\tau,1-2}$) are the total tangential electric (magnetic) fields on the two sides of the metasurface, respectively, and $Z_{s}$ denotes the corresponding surface impedance~\cite{Tretyakov_2003,Tretyakov_2017_homogenization,Christian_MT2018}. Here we assume that there is no cross-polarizing response. Since the tangential electric field is continuous across the sheet,  \mbox{$\_E\@{\tau,1}=\_E\@{\tau,2}=\_E\@{\tau}$}, and the difference between the  tangential magnetic field components is identical to the surface current density, \mbox{$\_n\times\left(\_H\@{\tau,2}-\_H\@{\tau,1}\right)=-\_J_{s}$} ($\_n$ is the unit vector orthogonal to the surface pointing into region~1), we can readily write that
\begin{equation}
\_n\times\left(\_H\@{\tau,2}-\_H\@{\tau,1}\right)=-\dfrac{\_E\@{\tau}}{Z_s}.
\label{eq:bound_cond}
\end{equation}

\begin{figure*}
\centering
\includegraphics[width=0.90\linewidth]{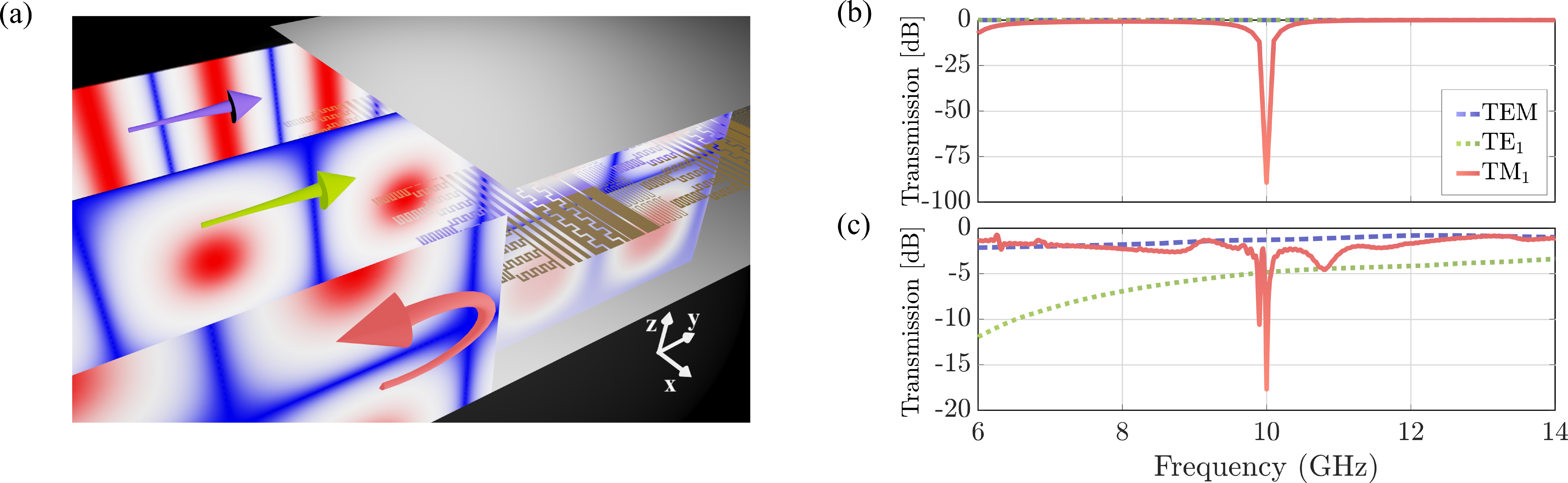}
\caption{\label{fig:stop_group} (a) A parallel-plate waveguide with allows propagation of TEM, TE$_1$ and TM$_1$ modes. However, a coherent retroreflective sheet placed between the plates acts as a mode filter, preventing propagation of a given mode (TM$_1$ mode in this proof-of-concept example). (b) and (c): Simulation results for a coherent retroreflective metasurface reveal that TM$_1$ can be suppressed at 10~GHz, without affecting propagation of other modes for (b) using six-discretized surface impedance stripes and (c) a metasurface implementation based on meandered slots and dipoles (see Summplementary Information).}
\end{figure*}

In order to function as a retroreflector, the metasurface requires different surface impedance profiles depending on whether it is illuminated with TE or TM waves.  In more detail, the desired field structure is a set of two incident plane waves, which are illuminating the opposite sides of the metasurface, and two retroreflected plane waves. Let us first consider the TE polarization, illustrated in Fig.~\ref{fig:teo_group}(a), in which case the fields are given by
\begin{equation}
\begin{split}
&\_E\@{I,1}=E\@{I,1} e^{-j k_0 \left(y {\rm sin} (\theta_i)-z {\rm cos} (\theta_i)\right)} \_x,\cr
&\_H\@{I,1}=-\dfrac{E\@{I,1}}{\eta_0}\Big[{\rm cos} (\theta_i)\_y+{\rm sin} (\theta_i)\_z\Big]e^{-j k_0 \left(y {\rm sin} (\theta_i)-z {\rm cos} (\theta_i)\right)},
\end{split}
\label{eq:TE_incw_1}
\end{equation}
\begin{equation}
\begin{split}
&\_E\@{R,1}=E\@{R,1} e^{j k_0 \left(y {\rm sin} (\theta_i)-z {\rm cos} (\theta_i)\right)} \_x,\cr
&\_H\@{R,1}=\dfrac{E\@{R,1}}{\eta_0}\Big[{\rm cos} (\theta_i)\_y+{\rm sin} (\theta_i)\_z\Big]e^{j k_0 \left(y {\rm sin} (\theta_i)-z {\rm cos} (\theta_i)\right)},
\end{split}
\label{eq:TE_reflc_1}
\end{equation}
in region 1, and 
\begin{equation}
\begin{split}
&\_E\@{I,2}=E\@{I,2} e^{-j k_0 \left(y {\rm sin} (\theta_i)+z {\rm cos} (\theta_i)\right)} \_x,\cr
&\_H\@{I,2}=\dfrac{E\@{I,2}}{\eta_0}\Big[{\rm cos} (\theta_i)\_y-{\rm sin} (\theta_i)\_z\Big]e^{-j k_0 \left(y {\rm sin} (\theta_i)+z {\rm cos} (\theta_i)\right)},
\end{split}
\label{eq:TE_incw_2}
\end{equation}
\begin{equation}
\begin{split}
&\_E\@{R,2}=E\@{R,2} e^{j k_0 \left(y {\rm sin} (\theta_i)+z {\rm cos} (\theta_i)\right)} \_x,\cr
&\_H\@{R,2}=\dfrac{E\@{R,2}}{\eta_0}\Big[-{\rm cos} (\theta_i)\_y+{\rm sin} (\theta_i)\_z\Big]e^{j k_0 \left(y {\rm sin} (\theta_i)+z {\rm cos} (\theta_i)\right)},
\end{split}
\label{eq:TE_reflc_2}
\end{equation}
in region 2. Here, $E\@{I}$ and $E\@{R}$ represent the amplitudes of the incident and reflected waves, respectively. Also, $\eta_0$ and $k_0$ are the free-space intrinsic impedance and wavenumber. We consider the full-reflection scenario, where all the incident power is reflected by the metasurface, with reflected fields defined as \mbox{$E\@{R,1}=E\@{I,1}\exp(j\phi\@{R,1})$} and \mbox{$E\@{R,2}=E\@{I,2}\exp(j\phi\@{R,2})$}. In addition, we suppose that the incident electric fields have the same magnitude, but there is a difference in phase: \mbox{$E\@{I,2}=E\@{I,1}\exp(j\phi\@{I,2})$}. Based on these considerations and  Eqs.~\eqref{eq:TE_incw_1}--\eqref{eq:TE_reflc_2}, we find the total tangential field components at the interface $z=0$ (see Supplementary Information), impose the boundary conditions, and satisfy Eq.~\eqref{eq:bound_cond}. This leads us to two important observations. First, total retroreflection is achieved only under coherent illumination with the  same phase ($\phi\@{I,2}=0$) and symmetric scattering ($\phi\@{R,1}=\phi\@{R,2}=\phi_R$). Second, the surface impedance must have the following profile:  
\begin{equation}
 Z_s^{\rm TE}(y) = j \frac{\eta_0 }{2\cos(\theta_i)}\cot\Big({\frac{\phi_R }{2}} + k_0 y\sin(\theta_i)\Big).\label{eq:imp_te}
\end{equation}
Thus, the surface impedance of coherent retroreflective sheets for TE polarization is a function of the position with the period $D=\lambda_0/2\sin(\theta_i)$ which depends on the angle of incidence and frequency (recall that $k_0=2\pi/\lambda_0$). It is worth mentioning that the effect of the reflection phase on the surface impedance is equivalent to a displacement of the retroreflective sheet along the $y$-axis, as portrayed in Fig.~\ref{fig:teo_group}(b). Simulation results, presented in Fig.~\ref{fig:teo_group}(c), demonstrate that total retroreflection is achieved  when the sheet is illuminated from both sides, while additional waves are scattered when it is illuminated from only one side.

The required surface impedance for transverse-magnetic (TM) polarization can be found in a similar way (see Supplementary Information), and it reads 
\begin{equation}
Z_s^{\rm TM} (y) = j\frac{\eta_0\cos(\theta_i) }{2} \cot\Big({\frac{\phi_R }{2}} + k_0 y\sin(\theta_i)\Big).
\label{eq:imp_tm}
\end{equation} 
As it is seen from the above two equations, surface impedances for TM and TE polarizations are quite similar, with the cosine term as the only difference.

\begin{figure*}
	\centering
	{\includegraphics[width=0.9\linewidth]{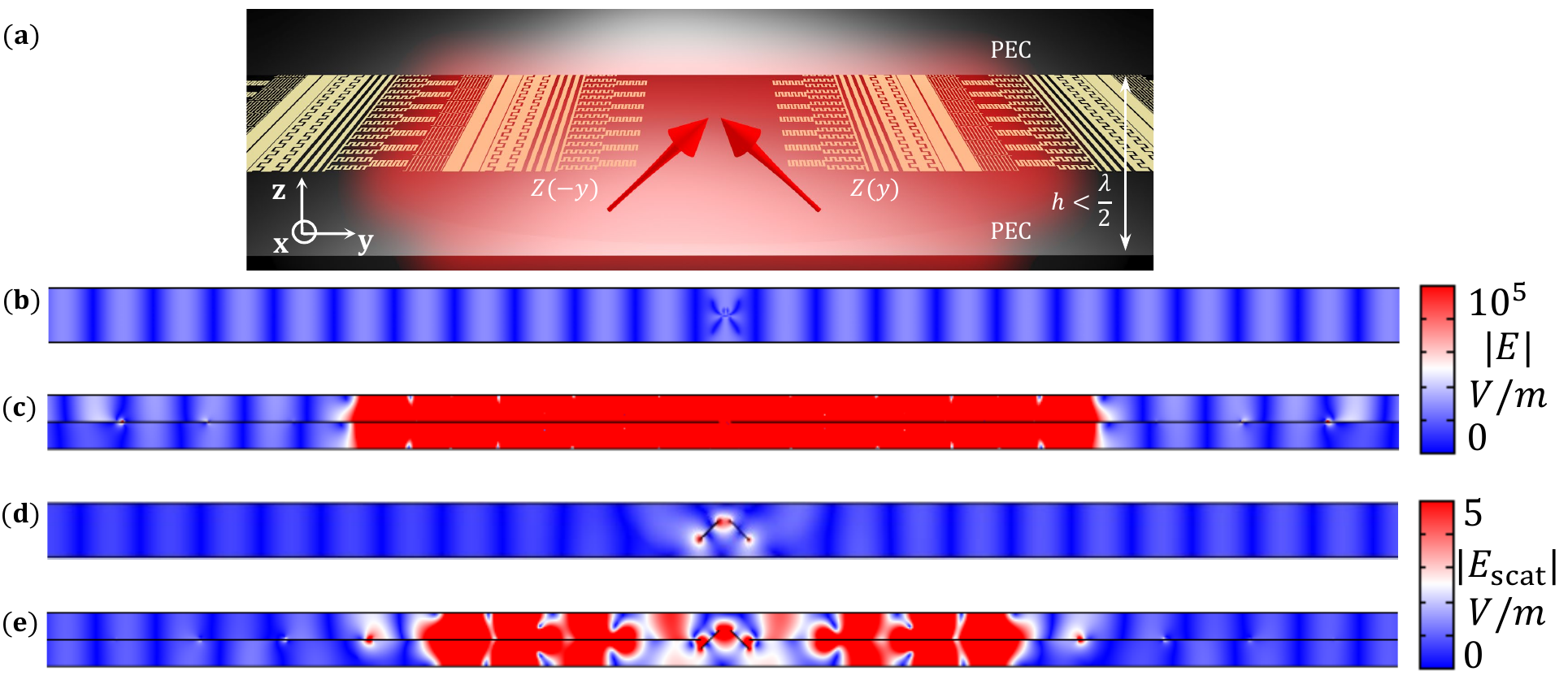}}
	\caption{(a) Schematic of a geometry using two retroreflectors for field localization. (b) and (c): Electric field amplitude created by two sources inside a waveguide without a metasurface in (b) and with metasurface in (c). (d) and (e): Scattered electric field amplitude of  waves created by a plane wave inside a waveguide without a metasurface in (d) and with metasurface in (e). The separation distance between the centers of wire scatterers is $\lambda_0/4$, and the operational frequency in all simulations is 10~GHz.}
	\label{fig:field_loc_group}
\end{figure*}



To illustrate the capabilities of surface-inhomogeneous coherent metasurfaces, we suggest two relevant applications where thin sheets can be effectively employed as coherent retroreflectors. The first application is about filtering undesired waveguide modes. We consider a parallel-plate waveguide which is formed by two ideally conducting sheets, as shown in Fig.~\ref{fig:stop_group}(a). The TM or TE modes propagating inside this waveguide can be expressed as superpositions of two plane waves having opposite components of the wave vector (the components that are perpendicular to the phase front)~(e.g.,~Ref.~\cite{Pozar_2004}).  Due to this crucial property, a coherent retroreflective metasurface, illuminated by these two plane waves, is able to prevent the propagation of a specific mode without disturbing the other modes. To realize this functionality, we need to position the metasurface in the middle of the waveguide, between the two boundaries, and choose the corresponding surface impedance properly. 

As a proof-of-concept example, let us assume that the waveguide has $h=33.31$~mm distance between the plates, enough to allow free propagation of transverse-electromagnetic (TEM), TE$_1$, and TM$_1$ modes at the reference frequency of $f_0=10$~GHz, even in the scenario when a thin PEC plate is placed in the middle of the waveguide. Suppose that the TM$_1$ mode is undesired, and, therefore, we are interested in blocking this specific mode. Since the angle of incidence for each mode is given by \mbox{$\theta_i=\arccos(n\lambda_0/(2h))$} in which $\lambda_0$ is the wavelength and $n$ represents the mode number~\cite{Pozar_2004}, we simply calculate the required surface impedance in accordance with Eq.~\eqref{eq:imp_tm} and the TM$_1$ mode. To see if such designed metasurface prevents the propagation of the undesired mode, we do numerical simulations using COMSOL Multiphysics software. Figure~\ref{fig:stop_group}(b) shows numerical results where the transmittance and reflectance of the three modes have been observed in the case when the designed coherent retroreflective metasurface, realized as six surface-impedance stripes \cite{Cuesta_2020_retroboundary} with a total length of $l=17.731$~mm (one period $D$), reference reflection phase $\phi_R=41.85^{\circ}$, and   incident angle $\theta_i=57.7157^{\circ}$, is placed in the middle of the waveguide. As it is seen, while the transmission of the TEM and TE$_1$ modes is not influenced by the presence of the metasurface over a large bandwidth, the TM$_1$ mode is effectively blocked by the metasurface at the design  frequency. This simulation was also performed for higher-order modes: TE$_2$ and TM$_2$, and we observed that the corresponding transmission of those modes remains unchanged with the presence of the metasurface, as expected. Similar results were obtained for the conceptual design shown in Fig.~\ref{fig:stop_group}(a) (see Supplementary Material for exact metasurface design\nocite{Cuesta_2020_retroboundary,Wang_2018_printable,Wang_2018_asymmetry}), where simulations carried in CST Microwave Studio show that the TM$_1$ mode was attenuated close to 20~dB at 10~GHz without affecting the other modes, as portrayed in Fig.~\ref{fig:stop_group}(c).

The second application is based on the use of coherent retroreflectors inside a waveguide to achieve significant field localization. A pair of coherent retroreflectors is aligned as shown in Fig.~\ref{fig:field_loc_group}(a). The surface impedance profile of one metasurface is mirrored with respect to the other one, so that both metasurfaces reflect back to the center of the structure. 
In this arrangement, some energy emitted by a small source positioned in the center is trapped  in the source vicinity as in a resonator, although there is no physical boundary in the waveguide. In order to demonstrate this concept, let us consider a narrow ($h<\lambda_0/2$) parallel-plate waveguide, such that only the TEM mode is supported. First, we position two Hertzian dipole sources oriented as shown in Fig.~\ref{fig:field_loc_group}(a) inside an empty waveguide, without the metasurfaces. In order to excite waves with a non-zero $y$-component of electric field, the dipoles are separated by a small distance $\delta\sim 0.025\lambda_0$. 
In the waveguide without retroreflective sheets, the dipoles excite TEM waves which propagate away from the sources, as illustrated in Fig.~\ref{fig:field_loc_group}(b). However, addition of two coherent retroreflectors creates a significant field-localization region, as is seen  in Fig.~\ref{fig:field_loc_group}(c). The result is obtained for the retroreflection phase $\phi_{\rm R}=343^{\circ}$.
Since the metasurfaces are infinitely thin, they do not interact with the TEM wave propagating along the $y$ direction, and we see that outside of the field-localization region the field is close to that of a single TEM mode. 

Similar field-localization effects can be achieved in the scattering regime. To demonstrate that, we remove the Hertzian dipole sources and position two small PEC (perfect electric conductor) wire scatterers at the center of the structure. The waveguide is excited by a wave port at the left end of the simulation domain. The two PEC wires are  rotated at the angle $\pm 45^{\circ}$ and separated by $\lambda_0/4$ distance, so that they excite enough strong  electric fields with the $y$ component in the vicinity of the scatterers  [see Fig.~\ref{fig:field_loc_group}(d)].
We need two wire scatterers in order to create two coherent scattered waves, equal in amplitude. Simulations confirm that placing metasurfaces around the wire scatterers enables us to trap significant portion of the scattered energy in the vicinity of the scatterers. Figure~\ref{fig:field_loc_group}(e) demonstrates localization of the scattered fields when the metasurface is inserted inside the waveguide. This result is achieved using the reflection phase $\phi_{\rm R}=305^{\circ}$. It is important to note that localization takes place without any physical boundary separating the volume of the ``virtual resonator'' from the rest of the waveguide. In these setups, the metasurfaces extends over the whole length of the waveguide. This property resembles bound states in the continuum, because the localization domain is fully open to both ends of the waveguide. The length of the localization domain is determined by the designed angle of retroreflection. 

In summary, we propounded the concept of coherent surface-inhomogeneous metasurfaces on an example of retroreflective sheets, whose attribute is to reflect plane waves back to the source when they are illuminated from both sides simultaneously. Unlike retroreflective boundaries, the proposed coherent retroreflective sheet reaches total retroreflection without the use of additional layers. This work presented scenarios of practical applications of non-homogeneous surfaces under coherent illumination. Through electromagnetic simulations, we have developed coherent retroreflective sheets that are capable to prevent propagation of a given electromagnetic mode in a multi-mode waveguide, without interfering with other modes; and to create localized field regions inside waveguides. The latter one can be viewed as a novel approach for realization of bound states in the radiation continuum. We expect that extending the concept of coherent inhomogeneous sheets to other metasurface functionalities will reveal other interesting new phenomena.




%

\end{document}